\documentclass[%
 reprint,
 amsmath,amssymb,
 aps,
]{revtex4-1}
\usepackage{graphicx}
\usepackage{dcolumn}
\usepackage{bm}

\newcommand{\be}{\begin{eqnarray}}
\newcommand{\ee}{\end{eqnarray}}
\newcommand{\rar}{\rightarrow}

\usepackage[]{caption}
\def\-g{\sqrt{-g}}

\renewcommand\rho{\varrho}

\let\jnlstyle=\rm
\def\refjnl#1{{\jnlstyle#1}}

\def\apj{\refjnl{Astrophys.~J.}}
\def\apjl{\refjnl{Astrophys.~J.~Lett.}}
\def\apjs{\refjnl{Astrophys.~J.~Supp.~Ser.}}

\def\mnras{\refjnl{Mon.Not.R.Astron.Soc.}}

\def\prc{\refjnl{Phys.~Rev.~C}}
\def\prd{\refjnl{Phys.~Rev.~D}}

\def\nat{\refjnl{Nature}}

\begin{document}
\title{Stars and Black Holes from the very Early Universe}

\author{A.D. Dolgov}
\email{dolgov@fe.infn.it}
\affiliation{Dipartimento di Fisica e Scienze della Terra, Universit\`a degli Studi di Ferrara\\
Polo Scientifico e Tecnologico - Edificio C, Via Saragat 1, 44122 Ferrara, Italy}
\affiliation{Istituto Nazionale di Fisica Nucleare (INFN), Sezione di Ferrara\\
Polo Scientifico e Tecnologico - Edificio C, Via Saragat 1, 44122 Ferrara, Italy}
\affiliation{Novosibirsk State University, Novosibirsk, 630090, Russia}
\affiliation{ A.I.~Alikhanov Institute for Theoretical and Experimental Physics, Moscow, 117218, Russia}
\author{S.I. Blinnikov}
\email{Sergei.Blinnikov@itep.ru}
\affiliation{ A.I.~Alikhanov Institute for Theoretical and Experimental Physics, Moscow, 117218, Russia}
\affiliation{Novosibirsk State University, Novosibirsk, 630090, Russia}
\affiliation{Sternberg Astronomical Institute, MSU, Moscow 119991; MIPT, Dolgoprudny, 141700,  Russia}
\begin{abstract}
A mechanism of creation of stellar-like objects in the very early universe, from the QCD phase transition 
till BBN and somewhat later, is studied. It is argued that in the considered process primordial black holes 
with masses above a few solar masses up to super-heavy ones could be created. This may explain an early 
quasar creation with evolved chemistry in surrounding medium and the low mass cutoff of the observed 
black holes. It is also shown that dense primordial stars can be created at the considered epoch. Such
stars could later become very early supernovae and in particular high redshift  gamma-bursters. In a version
of the model some of the created objects can consist of antimatter.
%
\end{abstract}

\maketitle

\section{ Introduction \label{s-intro}}

In the standard scenario of the cosmological evolution it goes without saying that the majority of
stars and galaxies have been formed very recently at redshifts of order unity, $z_{\rm form} \leq 10$.
Though the onset of star formation started much earlier at 
$z \approx 30$ for the so-called PopIII stars with zero metallicity, the fraction of baryonic matter in these early
formed stars is believed to be very low.
E.g. Tegmark et al. \cite{Tegmark} claimed that ``a fraction $10^{-3}$ of all baryons may have
formed luminous objects by $z = 30$''. 
Later Ricotti et al. \cite{Ricotti} have obtained that only $10^{-6}$ of all baryons are in stars at redshift
$z \sim 24 - 19$,
and the stellar fraction in baryons $10^{-3}$ is reached later, at $z \sim 15 - 14$.
Those numbers have been confirmed by Yoshida et al. \cite{Yoshida} and they are considered as a standard
for the star formation rate at the epoch of reionization. At the present time 
around 30\% of all baryons  are in stars and in intergalactic gas in clusters of galaxies.

The accepted history of the structure formation looks as
follows. At the very beginning, during inflationary stage, primordial density fluctuations with flat Harrison-Zeldovich
spectrum~\cite{harr,*zeld} were generated~\cite{mukh-chib}. These fluctuations remained practically frozen
during all radiation dominated (RD) epoch which came in place of inflation after the inflaton decay heated up the
universe. The RD epoch turned into the matter dominated (MD) one at redshift $z_{\rm eq} \sim 10^4$. After that initially small density
perturbations started to rise approximately as the cosmological scale factor,
$\Delta =\delta \rho /\rho \sim a(t)$. Since initially
$\Delta \sim 10^{-4}$, relative density fluctuations could reach unity to the present cosmological stage. 
When $\Delta$ became
large, the evolution of perturbations turns into a non-linear regime and they start to rise very quickly forming objects with
huge value of the ratio $\rho/\rho_c$, where $\rho_c= 10^{-29} $ g/cm$^3$ is the present day average cosmological energy
density. In this way stars, galaxies, and their clusters are believed to have been formed.

The essential time scales are the following. The universe age is $t_U = 13.8\pm 0.2 $ Gyr, according to the
recent Planck data~\cite{planck} with the Hubble constant $H_0 = 67.3 \pm 1.2$ km/s/Mpc, and a high 
value of the matter density parameter, $\Omega_m = 0.315 \pm 0.017$.
Galaxies and their clusters are supposed to be formed at $z = 2-3$, which correspond to the universe
age between 3.27-2.14 Gyr.

The lower limits on the ages of galaxies are determined by the ages of their stars. In particular, some stars
in the Milky Way are quite old with ages close to the universe age. For example, the age of  BD+17$^o$ 3248 was estimated
as $13.8\pm 4$ Gyr and a star in the galactic halo, HE 1523-0901, was estimated to be about 13.2 billion years old.
Moreover, recent observations indicate that the age of the star HD 140 283 is $14.46 \pm 0.31 $ Gyr~\cite{HD140}, whose
central value exceeds the universe age by two standard deviations.
Probably these stars are pregalactic ones formed independently of the galaxy and
captured by the galaxy much later. In the model, which is considered in this paper,
some stars or stellar-like objects could be
formed long before the galaxy formation epoch and behaved as cold dark matter.

On the other hand, there are several galaxies observed at high redshifts, with natural gravitational lens ``telescopes''.
In particular, there is a galaxy at $z \approx 9.6$ which was formed when the universe was approximately
0.5 Gyr old~\cite{gal-9.6},  and even a galaxy at $z \approx 11$, corresponding to the universe age 0.41 Gyr~\cite{gal-11}.

Another impressive example of early formed objects are quasars observed at high redshifts.
The maximum redshift of an observed quasar is 7.085, i.e. such quasar was formed when the universe was younger
than 0.75 Gyr.
The quasars are supposed to be supermassive black holes (BH) and their formation in such short time looks
problematic.
The models of an early formation of supermassive BHs are reviewed in
papers~\cite{BH-rev,*haiman,*VolonteriBellovary}.
For some recent references
see~\cite{BH-form,*DubrovichGlazyrin,*JohnsonWhalen,*WhalenJohnson,*Latif,*Choi,*Schleicher,*Gabor,*Muldrew,*Prieto}.
However, all the scenarios meet serious problems.
E.g., some scenarios \cite{BH-form} involve formation of very massive stars exploding as extremely powerful
supernovae.
Observations of very metal poor stars imply that their patterns of elemental
abundance are in good accord with the nucleosynthesis that occurs in stars with masses of
$(20-130)M_\odot$ when they become supernovae \cite{UmedaNomoto}. The abundances
are not consistent, however, with heavy element enrichment by supernovae originated from
more massive stars in the range (130-300) M$_\odot$.
It is inferred \cite{UmedaNomoto} that
the first-generation supernovae came mostly from explosions of
$\sim (20-130) M_\odot$ stars.

There are strong indications that every large galaxy, as well as some relatively small ones~\cite{BH-small-gal},
contains central supermassive black hole.
The mass of the black hole may be larger than ten billions $M_\odot$ in giant elliptical
and compact lenticular galaxies
and about a few million $M_\odot$ in spiral galaxies like Milky Way (MW).
The mass of the BH
in the MW center is about  $\sim 10^{-5}$ relative to the total MW mass. Normally, the BH mass
is smaller in spiral galaxies and is correlated with the bulge mass, but not with the total mass of the 
galaxy~\cite{BH-morph,*Zasov05,*Zasov11,*McConnellMa}.
(MW has a BH which lies below the value determined by this correlation, perhaps this is good: 
otherwise, the life on the Earth could be threatened by the quasar radiation).

The mass of the
black hole is typically 0.1\% of the mass of the stellar bulge of
the galaxy \cite{BH-bulge,*SaniMarconi}
while some galaxies may  have a huge BH: e.g. NGC 1277  has
the central black hole  of  $1.7 \times 10^{10} M_\odot$, or $\sim  60$\% of its bulge mass \cite{NGC1277}.
This fact creates serious problems for the
standard scenario of formation of central supermassive BHs by accretion of matter in the central part of a galaxy.
An inverted picture looks more plausible, when first a supermassive black hole was formed and it attracted matter
serving as a seed for subsequent galaxy formation.
The mechanism of such early BH formation is discussed below.

It is striking that the medium in the vicinity of such early quasars contains
considerable amount of ``metals'' (i.e. of
elements heavier than helium), see e.g. ref.~\cite{QSO-chemistry}.
According to the standard picture, only elements up to $^4$He  { and traces of Li, Be, B}
were formed in the early
universe during Big Bang Nucleosynthesis (BBN), while heavier elements were created
by stellar nucleosynthesis and dispersed in the interstellar space by the supernova explosions.
It means that according to the standard scenario
prior to creation of quasars, an efficient star formation processes should take place in the universe.
These stars evolved producing supernovae and the latter enriched the space around them  by metals.

The duration of pre-supernova stellar evolution is about 13
Myr for the stars with the  initial mass $15 M_\odot$ and 3.5 Myr for those with with the initial mass
$75 M_\odot$~\cite{StarEvolution}.
The values of lifetime for the ordinary stars are given from their
formation until supernova explosion (or collapse to a BH, cf. \cite{StarEvolution,LimongiChieffi}).
But the ordinary stars are composed by 70\% of hydrogen, while the new types of stars, considered here,
are initially almost pure helium, since they came out matter where BBN proceeded with much larger
baryonic density than the standard one. (We call such stars the AD-stars by the reason explained below.)

Nevertheless, separate calculations of stellar evolution for the AD-stars are not needed:
each ordinary massive star, after hydrogen is burnt out in the central regions,
has a helium core, which quickly reaches half the mass of the original star
(with high accuracy $\sim10$\% \cite{StarEvolution}).
The remarkable fact is that this helium core lives independently of
the amount of hydrogen left in the envelope (moreover, almost all of the  hydrogen
in the outer layers of a red supergiant  may be lost 
in the stellar wind \cite{LimongiChieffi}, and we are left with a bare helium star, the so called Wolf-Rayet star).
Therefore, for an AD-star, we can take the existing calculations for the evolution of the normal
stars, and extract out of them the lifetime on the stage of the helium core.
That is, if we need to find the lifetime of a 10~$M_\odot$-AD-star, we take an ordinary star
of 20~$M_\odot$ and find its lifetime on the helium burning stage equal to 1.2 Myr.
Thus, the ordinary star with $M=15\,M_\odot$ corresponds to a He-star of about 7 $M_\odot$, and an ordinary
star of 75 $M_\odot$ corresponds to a He-star of about 30 $M_\odot$.
The lifetimes of the two AD-stars with masses 7 and 30 $M_\odot$ are respectively 
2 Myr and 0.5 Myr~\cite{StarEvolution}.
For  $M > 75 M_\odot$ the lifetime becomes almost independent of $M$ since the luminosity is close 
to the Eddington limit and thus is proportional to $M$  as is also true for the nuclear energy supply.

Observations of high redshift gamma ray bursters (GRB) also indicate a high abundance of supernova at
large redshifts.  The highest redshift of the observed GRB is 9.4~\cite{GRB-max} and there are a few more
GRBs with smaller but still high redshifts. The necessary star formation rate for explanation of these early
GRBs is at odds with the canonical star formation theory.

A recent discovery of an ultra-compact dwarf galaxy~\cite{ultra-dwarf} older than 10 Gyr, 
enriched with metals, and probably with a massive black in its center seems to be at odds with 
the standard model but may well fit the scenario discussed in this paper. 

\section{Early formation of stellar-like objects}

In this note we consider a model of the efficient formation of stellar-like objects in
the very early universe which seems to
naturally resolve the above mentioned problems. The model was suggested in paper~\cite{ad-silk} and further
refined in ref.~\cite{ad-mk-nk}. The considered scenario is based on a slightly modified Affleck-Dine (AD)
suggestion for the  baryogenensis~\cite{affleck-dine}
where the general renormalizable coupling of the scalar baryon, $\chi$, to the inflaton field, $\Phi$,  is introduced:
\be
U(\chi, \Phi) = U_\chi (\chi) + U_\Phi (\Phi) + U_{\rm int} (\chi,\Phi).
\label{U-of-Phi-chi}
\ee
Here $ U_\Phi (\Phi) $ is the inflaton potential, depending upon the model of inflation,  $ U_\chi (\chi)$ is
the quartic Affleck-Dine
potential, which has some flat directions (valleys), and  the additional interaction term has the form:
\be
U_{\rm int} (\chi,\Phi) = \lambda_1 |\chi|^2 \left( \Phi - \Phi_1\right)^2 ,
\label{U-int}
\ee
where $\Phi_1$ is some value of the inflaton field which it passes during inflation and $\lambda_1$ is a
constant (we keep the notations of ref.~\cite{ad-mk-nk}).

The baryogenesis in AD scenario proceeds as follows. At inflationary stage field $\chi$ may reach large
values due to rising quantum fluctuations along the flat directions of $U_\chi$. When inflation is over,
$\chi$ evolves down to the minimum of
the potential, which is supposed to be at $\chi = 0$. On the way down $\chi$ acquires some ``angular momentum'' in the
complex plane $[ \mbox{Re}\, \chi, \mbox{Im}\, \chi]$. This happens either due to quantum fluctuations in the direction
orthogonal to  the valley
or because of mismatch of the flat directions of $\chi^4$ and $\chi^2$ terms in the potential $U_\chi$.
This ``angular momentum'' is proportional
to the baryonic charge of $\chi$: $B_\chi \sim i  [(\partial_0 \chi^*) \chi - \chi^* \partial_0 \chi)]$.
It is released later into baryonic charge of quarks in the process of $B$-conserving decay of $\chi$.
This process could lead to a huge cosmological baryon asymmetry, $\beta = N_B /N_\gamma $,
which might be close to unity, i.e. much larger than the observed canonical value,
$\beta \approx 6\cdot 10^{-10}$.

An addition of $U_{\rm int}$-term into $U(\chi,\Phi)$, eq.~(\ref{U-of-Phi-chi}), strongly changes the evolution
of $\chi$. When
$\Phi \neq \Phi_1$, the effective mass of $\chi$ is positive, so the gates to the valleys are closed and
most probably $\chi$ rests near $\chi  = 0$. Hence the baryogenesis in most of the space proceeds with normal
low efficiency producing the observed small value of $\beta$. However during the time when the gates
to the valley are open, i.e. when $\Phi$ is close to  $\Phi_1$, the baryonic scalar $\chi$ may ``rush'' to
large values. The probability of this process
is low and so the bubble with large baryonic asymmetry would occupy a small fraction of space forming some compact
objects with a large baryonic number. The details and (numerical) calculations can be found in ref.~\cite{ad-mk-nk}.

The perturbations initially induced by such process are predominantly isocurvature ones, i.e. they have
large variation of the
baryonic number, $\delta B / B \gg 1$ with small perturbations in the energy density, $\delta \rho/\rho \ll 1$.
Situation drastically changes after the QCD phase transition at the cosmological temperature of about 100 MeV.
After that
practically massless quarks turn into massive baryons and excessive baryonic number contained in high $B$ bubbles
leads to large density contrast between these bubbles and low-$B$ background. There appear compact objects with
log-normal mass distribution:
\be
\frac{dN}{dM} = C_M \exp\left[-\gamma\,\ln^2(M/M_0) \right]
\label{dN-dM}
\ee
where $C_M$, $\gamma$, and $M_0$ are some constant parameters. The shape of the distribution is practically
model independent. It is determined by the exponential law of expansion during inflation, but the
values of the parameters are model dependent and thus not  known.
We take them as free parameters in a range which seems reasonable.

If $\delta \rho /\rho$ in such bubbles happened to be larger than unity at the horizon scale, then they
would form primordial black holes (PBH) created at first seconds or even at a fraction of second of the universe
life. If $\delta \rho /\rho <1$ at the horizon
crossing, then PBH would not be formed but instead some stellar-like objects would be created at this early time.
The value of $\delta \rho /\rho$ at horizon depends upon the magnitude of $\beta$, which is not a constant
but more or less
uniformly distributed quantity for different bubbles. It is worth noting that $\beta$ may be even negative thus resulting in
a noticeable amount of antimatter in the form of compact objects in the Galaxy. For phenomenology of such
antimatter objects see ref.~\cite{bambi-ad}.

The mass distribution (\ref{dN-dM}) naturally explains some of the observed features of the distribution
of stellar mass black holes in the Galaxy.
For example in ref.~\cite{BH-masses-1} it was found that the masses of the observed black holes
are best described by a narrow mass distribution at  $ (7.8 \pm 1.2) M_\odot $.
This result agrees with ref.~\cite{BH-masses-2,*FarrSravan} where
a peak around $8M_\odot$, a paucity of sources with masses below $5M_\odot$, and a sharp drop-off above
$10M_\odot$ are observed. These features are not explained in the standard model.

Moreover,  simple modifications of the  interaction potential (\ref{U-int}) would lead to a more 
interesting/complicated mass spectrum of the created black holes and other early formed stellar type objects. 
For example, taking $U_{\rm int} $ in the form:
\be
U_{\rm int} (\chi,\Phi) = \frac{\lambda_1}{M_2^2} |\chi|^2\,  \left( \Phi - \Phi_1\right)^2   \left( \Phi - \Phi_2\right)^2 ,
\label{U-int-2}
\ee
we come to a two-peak mass distribution of these primordial black holes, which is 
observed, see papers~\cite{BH-masses-1,BH-masses-2,*FarrSravan} 
and the references therein, but not explained up to
now \footnote{We thank V.Sokolov for informing us about the data on the two-peak
mass distribution of black holes.}.

Evolved chemistry in the so early formed QSOs can be also explained, at least to some extend,
by more efficient production of
metals during BBN due to much larger value of the baryon-to-photon 
ratio $\beta =N_B/ N_\gamma$. In the standard cosmology BBN essentially stops at $^4$He due to very
small $\beta$. However, as we have mentioned above, in the model considered here $\beta$ may be
much larger than the canonical value, even being close or exceeding unity. BBN with high $\beta$ was
considered in ref.~\cite{BBN-hi-beta,*Matsuura05,*Matsuura07,*MatsuuraDolgovNagataki},
where it was shown that the outcome of metals is
noticeably enhanced, though the calculations have been done only for moderately large $\beta$, not larger than
than 0.001, which is ``only'' 6 orders of magnitude larger than the standard baryon asymmetry.
The latter, in terms
of the present day ratio of baryon-to-photon number densities, is equal to:
\be
\beta_0 = 6 \cdot 10^{-10}.
\label{beta-0}
\ee
The predictions of the standard BBN are not distorted because the unusual abundances of light elements
are concentrated only in a tiny fraction of space and their diffusion out is very short.

Depending upon the value of the baryon-to-photon ratio, $\beta_B$, inside the bubbles and the bubble size, $R_B$,
such high baryon density objects could form either a primordial black hole (PBH), or a kind of star, or a disperse cloud
of gas with unusually high baryonic number. The selection between these  possibilities depends upon the
value of the Jeans mass of the objects.

It is convenient to specify the initial conditions at the moment of the QCD phase transition (p.t.) in the primeval plasma,
after which massless free quarks turned into heavy baryons,
i.e. into protons and neutrons, with $m \approx 1 $ GeV. After such
p.t. the (quasi)isocurvature density perturbations initially with $\delta \rho \approx 0$ led to the density contrast
$\delta \rho = \beta_B N_\gamma m $, if densities (and temperatures) of photons inside and outside the bubbles
are assumed to be the same. The relative density contrast is equal to
\be
\delta \rho /\rho_c \approx 0.2 \beta_B (m/T),
\label{delta-rho-over-rho}
\ee
where $\rho_c = 3 H^2 m_{Pl}^2 / (8\pi) $ is the cosmological energy density and $\beta$ is normalized to  the
present day values of baryon and photons densities, where the heating of the photons by $e^+e^-$-annihilation
is taken into account, while $N_B$ is supposed to be conserved in the comoving volume and the baryon diffusion
out of the bubble is neglected.

At the QCD p.t. the universe is dominated by the equilibrium relativistic matter with temperature
$T$, so $H=1/(2t) $  and the cosmological energy density is
\be
\rho_c =  \frac{3 H^2 m_{Pl}^2 }{ 8\pi} = \frac{\pi^2 g_* T^4}{30} ,
\label{rho-c}
\ee
where $g_*$ is the number of the
relativistic degrees of freedom. The temperature of the QCD p.t., $T_Q$, is not well known. It is somewhere in the
interval $ T_{QCD} =100-200$ MeV. In this temperature interval but after p.t. $g_* =  17.25$, while below $100$ MeV:
$g_* = 10.75$. Thus the relation between the cosmological time and temperature is
\be
t/{\rm sec} = 0.7\cdot 10^{-4} \left( \frac{10.75}{g_*} \right)^{1/2}
  \left( \frac{100\,{\rm  MeV}}{T}\right)^2.
\label{t-of-T}
\ee
The mass inside horizon,  $l_h = 2 t$, is equal to
\be
M_h = m_{Pl}^2 t = 10^5 M_\odot (t/{\rm sec}) = \nonumber\\
14 M_\odot  \left( \frac{10.75}{g_*} \right)^{1/2} \left( \frac{100\,{\rm  MeV}}{T}\right)^2.
\label{M-h}
\ee

We denote the universe age, $t$, the temperature, $T$, and the radius of the bubbles, $R_B$ at the moment
of the QCD p.t.  as $t_Q$, $T_Q$, and $R_Q$ respectively. The radius is stochastically distributed quantity,
whose distribution is analogous to (\ref{dN-dM}). The baryon asymmetry inside the bubbles, $\beta$, is also a
stochastic quantity, which we assume to be uniformly distributed between $\beta_{max}$ and $\beta_{min}$.

The bubble would form a PBH at horizon crossing if its radius is
smaller than the gravitational radius of the bubble, $r_g = 2 M_B/ m_{Pl}^2$, where the mass of the bubble is equal to
\be
M_B = \frac{4 \pi}{3}\, R_B^3 \rho_B = \frac{4\pi^3 g_*}{90} R^3_B T^4 (0.2 \beta m/T).
\label{M-B}
\ee
So the condition of PBH formation is
\be
0.2 \beta\, \frac{m}{T}\, \left(\frac{R_B}{2t}\right)^2 >1.
\label{PBH-form}
\ee
Thus if $\beta \sim 1$ the bubble would become a PBH at the QCD p.t. if $R_Q /(2t_Q ) =  1$. If $\beta_{max} = 1$, then
the smallest mass of PBHs formed in this way would be equal to the mass inside horizon at $t=t_Q$. Taking $T_{QCD} = 150$ MeV,
we find that the PBH mass should be above $5 M_\odot$, which is very close to the upper limit below which black holes
are not observed~\cite{BH-masses-1,BH-masses-2}. No other explanation for this cut-off has been found.

If $\beta > 1$, then PBH formation with smaller masses, i.e. corresponding to $R_Q/(2t_Q) <1$, is  also possible at the QCD
p.t. In this case PBHs would be formed practically instantly, when massless quarks turned into massive baryons and the density
contrast jumped from zero to that given by eq. (\ref{delta-rho-over-rho}). For PBH formation
the condition $ \beta > 5 (T_Q/m) (2t_Q/R_{BQ})^2$ should be fulfilled, as one can see
from eq.~(\ref{PBH-form}).
According to a simple version of  the model~\cite{ad-silk,ad-mk-nk}, very large $\beta$  is unlikely,  though not excluded,
and the formation probability  of lighter PBHs is most probably small.

Heavier PBHs, which could be formed in the considered scenario,  originated from the bubbles whose radius
was larger than horizon at QCD p.t., $R_Q /(2t_Q) > 1$.  As we have mentioned above in slightly different terms,
PBHs would be created if at the horizon crossing $\delta \rho /\rho > 1$.
Assuming that this took place at the RD stage
when the cosmological scale factor rose as $a(t) = a_Q (t/t_Q)^{1/2} $,
temperature dropped as $T =T_Q (a_Q/a)$, and the bubble expanded together with the universe, i.e.
$R_B(t) = R_Q\, a(t) /a_Q$, we find that the moment of the horizon crossing, $t_h$, is given by
\be
t_h = R_Q^2 / 4 t_Q
\label{t-h}
\ee
The corresponding temperature is $T_h = T_Q (t_Q/ t_h)^{1/2}$ and we find that PBH would be formed if
\be
0.2\beta\,\frac{m}{T_Q}\,\frac{R_Q}{2t_Q} > 1
\label{PBH-form-low-beta}
\ee
This condition is not precise. It may happen that $\delta\rho /\rho$ reached unity before the horizon crossing
and the rise of $R_B(t)$ would slow down, but for the moment we neglect these subtleties.

Note the difference between conditions (\ref{PBH-form}) and (\ref{PBH-form-low-beta}). It reflects the difference of
physics in PBH formation. In the first case the PBH is formed when the density inside a small bubble with $R_B < l_h$,
suddenly rose up and the  bubble collapsed, while the second case is the usual story of PBH creation in cosmology.
As one should expect, conditions (\ref{PBH-form}) and (\ref{PBH-form-low-beta})  coincide at $R_Q = 2t_Q$.
However, our approach is oversimplified and the formation of PBH with $R_Q < 2t_Q$ at QCD p.t. may be much more
complicated process, when the rise of the energy density inside the bubble and
the effects of general relativity could terminate or postpone the phase transition.
The problem of the bubble formation at phase transitions and in particular of black holes
was studied in ref.~\cite{Berezin83,*Berezin87}.

Those bubbles which avoided becoming PBHs, formed all kinds of compact stellar-like objects
or much lower density  clouds. The evolution of such objects created in the very early universe depends upon
the ratio of the bubble mass to their Jeans mass. We can call such stellar-like objects either
BB-stars (baryonic bubble stars) or AD-stars, because they could be created as a result of Affleck-Dime
baryogenesis. Their properties can be quite different from the normal stars at their initial stage.
For example the initial temperature inside the bubble could be
smaller than the temperature of the cosmological matter outside because nonrelativistic matter cools faster during
cosmological expansion. Correspondingly the external pressure would be larger than the internal one.
Later when the bubble decoupled from the expansion and started to shrink due to its own gravity,
its temperature gradually became
larger than the outside temperature and the situation would be closer to the normal astrophysics.

The mass of the created AD-stars is roughly equal to the mass inside their radius, $R_Q$, at the
QCD phase transition:
\be
M_{\rm AD} = \frac{4\pi  R^3_Q  }{3}  \frac{ \rho_Q  \delta \rho_Q}{\rho_{c Q}} =
\xi^3 \beta \, (m_{Pl}^2 t_Q ) \, \frac{0.2 m }{T_Q}  ,
\label{M-AD}
\ee
where $m_{Pl}^2 t_Q \approx 3.5 M_\odot\, (200\,\, {\rm MeV} /T_Q)^2$
is the mass inside horizon (\ref{M-h})  at the
QCD p.t. for the average cosmological energy density  (\ref{rho-c}), $\xi \equiv R_Q/(2t_Q)$, 
and the relative density contrast is given by
eq. (\ref{delta-rho-over-rho}). Taking for simplicity
$ 2 m/T_Q = 1$, i.e. $T_Q = 200$ MeV, we find that the mass of AD-star is $M_{\rm AD} = 3.5 M_\odot\xi^3 \beta$,
the temperature when $\delta \rho /\rho_c = 1$ is
$T_1 = 0.2 \beta m$, and the condition to avoid becoming a PBH is $\beta \xi <1$.

Let us consider as an example a bubble with the mass close to the solar one and $T_1 \sim 50 $ keV. The energy
density at the moment, when $\delta\rho = \rho$ would be about $10^8$ g/cm$^3$.
The thermal energy of a solar mass B-bubble taken at the moment
when the Jeans mass dropped down to $M_\odot$ is determined by the
thermal energy of nucleons, $E_{th} = 3T/2$ (electrons are degenerate at those densities).
Taking  $T= 50 $ keV, though the temperature may
drop down due to AD-star initial expansion, we find for the total energy
stored inside this "star":
\be
E_{\rm therm}^{\rm(tot)} = \frac{3 T M_{\rm AD}}{  2m_N }\approx   10^{29} {\rm g}
\approx 10^{50} {\rm erg} .
\label{E-therm}
\ee

In the considered example with $\rho \sim 10^8$ g/cm$^3$
AD-star has the properties similar to those of the core of  a red giant
at the initial stage of its evolution.  The main source of energy under
these conditions would be helium-4 burning, $3\,^4\mbox{He} \rar ^{12}\mbox{C}$.
However, in the considered example the temperature, $T\sim$ 50 keV,  is noticeably
larger than that of  the normal red giant core, $T_{\rm rg} \sim 10$ keV.
Since the probability of the above reaction is a strong exponential function of $T$,
its rate at  $T\sim$ 50 keV is 10 orders of magnitude higher than at $T_{\rm rg}$
\cite{Ishikawa}.
The life-time of such helium flash in the AD-star would be extremely short. Naively taking
these numbers, we obtain life-time about a few hours instead of million years discussed
in Sec.\ref{s-intro} for He-stars. However, this simple estimate
can be wrong by several orders of magnitude because the efficiency of the
process is very much different from that in normal giant star. 
{Since the hydrodynamic time is $\sim {G_N \rho}^{-1/2}$, i.e. less than a second, the initial
B-ball would expand and cool down quickly to a normal $T_{\rm rg}$ well before He is exhausted.
Thus an AD-star would be formed with the properties similar to normal He-stars.}
Still a fraction of helium would be burnt very quickly at 
the very beginning and other nuclear reactions, which could occur later, would be presumably 
insignificant for the full life-time of the star, since later nuclear reactions are even faster.
More accurate estimates would demand development
of astrophysics of such strange objects as B-balls. One needs to study evolution
of many unusual types of prestellar objects which may be very much different  from the
standard stellar evolution, at least at the initial stage.

\section{Discussion}

The main presently observable cosmological impact of AD-stars  is
the enrichment of the interstellar space by metals which was a result of their fast 
evolution and subsequent explosion in distant past. In addition, as a result of their
evolution there could be formed peculiar stars of huge age made of ordinary matter, early black
holes, and gamma-bursters which are observed today. Moreover, AD-stars could
give birth to old low mass cold helium red dwarfs, dead white dwarfs, and neutron stars.

Normal single stars may either evolve to
core-collapse  at the mass of He core
$2 M_\odot \lesssim M_{\rm He} \lesssim 40 M_\odot$ or to pair-instability supernovae at
$M_{\rm He} > 40 M_\odot$ \cite{StarEvolution}.
The life-time of a massive star with $M_{\rm He} > 40 M_\odot$ is less than
1 Myr during the stage of He burning \cite{StarEvolution}.

Such a massive star can produce a good supernova within a Myr after recombination. 
With $\rho_c = 10^{-29}$ g/cm$^3$ and $\Omega_b = 0.05$ we would have the present 
day average cosmological
mass density of baryons equal to $\rho_b = 5\cdot 10^{-31}$ g/cm$^3$. At recombination it would be 9 orders
of magnitude higher, i.e. $\rho_b = 5\cdot 10^{-22}$ g/cm$^3$ or about 100 baryons per cubic cm.
If the AD-star lives a bit less than a Myr then at the moment of its supernova explosion
the cosmological density of the environment in ``our'' ordinary baryons,
even ignoring the growth of perturbations, is still high, of the order of a few  baryons per cubic centimeter.
In other words, it is the same as the present day density in the dense regions of gaseous disk of our Galaxy.
That is, AD-supernova explosion occurs in an environment that we understand reasonably well,
except for the fact that the interstellar medium had  a different chemical composition.
Even if we do not understand all the details theoretically, we observe the metal-enriched composition of the
interstellar medium, coming  presumably from the remnants of such
explosions i.e. from the Supernova Remnants  (SNRs).

However, in the case of AD-stars their own chemical composition should also be contaminated with
metals due to the non-standard BBN as well as the chemical composition of the interstellar medium,
due to the stellar wind and the AD-supernova explosion.
We observe that ordinary SNRs are associated with regions of star formation. After all,
a few tens thousand years after the explosion, the uniform interstellar medium would be
swept up into a thin wall of the SNR-bubble with a mass of thousands of solar masses.
With sufficient abundance of metals it would be catastrophically cooled down
generating thousands of young stars.
Supernova remnants do not produce very massive star, but they naturally give birth to small
ones, with  masses around $1 M_\odot$ and less,
just as it is necessary for the ``prehistoric'' star HD 140283.

Thus the described scenario
leads to very interesting consequences, such as formation of stellar mass PBHs, as well as of
supermassive BHs and the first supernovae which could lead to formation of peculiar stars like HD 140283.
This helps to resolve the problems of the early formation of black holes,
quasars, GRBs, as well as all the first stars, and the enrichment by metals
of the  interstellar space at high redshifts.
At the tail of the distribution (\ref{dN-dM}) supermassive PBHs could be created which might serve as seeds
for galaxy formation. Another interesting and testable consequence of the discussed scenario is prediction of
compact stellar type objects made of antimatter which might abundantly populate the halo of the Galaxy.

\begin{acknowledgements}
We are grateful to Vladimir Sokolov for stimulating comments.
We acknowledge the support by the grant of the Russian Federation government 11.G34.31.0047.
The work of SB is also partly supported by grants for 
Scientific Schools 5440.2012.2, 3205.2012.2, and joint RFBR-JSPS grant 13--02--92119.
\end{acknowledgements}

\providecommand{\noopsort}[1]{}\providecommand{\singleletter}[1]{#1}%
%


\end{document}